\documentclass[prd,preprint,superscriptaddress,tightenlines,nofootinbib,eqsecnum,showpacs]{revtex4-1}

\usepackage{amsmath}
\usepackage{amsfonts}
\usepackage{amssymb}
\usepackage{mathrsfs}
\usepackage{bm}
\usepackage{hyperref}
\usepackage{color}

\definecolor{CiteColor}{rgb}{0,0,0.35}
\definecolor{URLColor}{rgb}{0,0,0.35}
\hypersetup{colorlinks,citecolor=CiteColor,urlcolor=URLColor}

\newcommand{\beq}{\begin{equation}}
\newcommand{\eeq}{\end{equation}}
\newcommand{\ud}{\mathrm{d}}
\newcommand{\ui}{\mathrm{i}} 
\newcommand{\ue}{\mathrm{e}} 
\newcommand\calO{{\mathcal{O}}}
\newcommand\scrR{{\mathscr{R}}}
\newcommand{\bo}[1]{\mathbf{#1}}
\newcommand{\av}[1]{\langle #1 \rangle}

\allowdisplaybreaks

\begin{document}

\title{First Law of Compact Binary Mechanics \\ with Gravitational-Wave Tails}

\author{Luc \textsc{Blanchet}}\email{blanchet@iap.fr}
\affiliation{$\mathcal{G}\mathbb{R}\varepsilon{\mathbb{C}}\mathcal{O}$,
  Institut d'Astrophysique de Paris, UMR 7095, CNRS, Sorbonne
  Universit{\'e}s \& UPMC Univ Paris 6, 75014 Paris, France}

\author{Alexandre \textsc{Le~Tiec}}\email{letiec@obspm.fr}
\affiliation{LUTH, Observatoire de Paris, PSL Research University, CNRS,
Universit\'e Paris Diderot, Sorbonne Paris Cit\'e, 92190 Meudon, France}

\date{\today}

\begin{abstract}
We derive the first law of binary point-particle mechanics for generic
bound (i.e. eccentric) orbits at the fourth post-Newtonian (4PN) order,
accounting for the non-locality in time of the dynamics due
to the occurence of a gravitational-wave tail effect at that
order. Using this first law, we show how the periastron advance of the
binary system can be related to the averaged redshift of one of the
two bodies for a slightly non-circular orbit, in the limit where
the eccentricity vanishes. Combining this expression with
existing analytical self-force results for the averaged redshift, we
recover the known 4PN expression for the circular-orbit periastron
advance, to linear order in the mass ratio.
\end{abstract}


\maketitle

\section{Introduction} 
\label{sec:intro}

Analytic approximation methods in General Relativity, such as the
post-Newtonian (PN) approximation \cite{Sc.11,Bl.14,FoSt.14,Po.16},
gravitational self-force (GSF) theory \cite{Po.al.11,Ba.14,Po2.15},
and the effective one-body (EOB) model \cite{Da.14}, play an important
role both in the data analysis of gravitational waves, and for
comparisons with the results from numerical relativity (NR)
simulations \cite{Le2.14,BuSa.15}. Recently, significant progress has
been achieved on the derivation of the equations of motion of binary
systems of compact objects at the fourth post-Newtonian (4PN) order,
using the Arnowitt-Deser-Misner (ADM) canonical Hamiltonian formalism
in ADM-TT coordinates
\cite{JaSc.12,JaSc.13,JaSc.15,Da.al.14,Da.al.16}, the Fokker action
approach in harmonic coordinates \cite{Be.al.16,Be.al.17}, and
effective field theory (EFT) methods
\cite{FoSt.13,Ga.al.16,Fo.al.17}. The next objective is to compute the
gravitational radiation field at the 4PN order (beyond the lowest
order quadrupole radiation). So far, specific high-order tail effects
in the waveform and energy flux have already been computed
\cite{Ma.al.16}.

Gravitational wave tails also form an integral part of the
conservative dynamics starting at the 4PN order (beyond the Newtonian
motion) \cite{BlDa.88,Bl.93,FoSt2.13,Ga.al.16}. They bring on an
interesting new feature of the conservative two-body dynamics at this
order of approximation, namely the \textit{non-locality} in time.
This has been shown in the canonical ADM Hamiltonian
\cite{JaSc.12,JaSc.13,JaSc.15,Da.al.14,Da.al.16}, the
harmonic-coordinates Fokker Lagrangian \cite{Be.al.16,Be.al.17} and
the EFT \cite{FoSt2.13,Ga.al.16} approaches. The 4PN tail term is
related to the appearance of infra-red divergencies in the ADM and
Fokker formalisms, and such divergencies have entailed the presence of
``ambiguity parameters'' that have plagued---for the moment---the
derivations of the 4PN dynamics
\cite{JaSc.12,JaSc.13,JaSc.15,Da.al.14,Da.al.16,Be.al.16,Be.al.17,PoRo.17}.

On the other hand, the conservative dynamics of binary systems of
compact objects enjoys a fundamental property now known as the
\textit{first law of binary mechanics}. For circular orbits, this law
is a particular case of a more general variational relationship, valid
for systems of black holes and extended matter sources
\cite{Fr.al.02}. The first law for non-spinning point-particle
binaries on circular orbits was established in
Ref.~\cite{Le.al.12}. It was later generalized to spinning binaries
\cite{Bl.al.13}, and more recently extended to generic bound
(eccentric) orbits \cite{Le.15}. These laws have been derived on
general grounds, assuming that the conservative dynamics of the binary
derives from an autonomous canonical Hamiltonian.
Moreover, they have been explicitly checked to hold true
up to 3PN order, and even up to 5PN order for some logarithmic
terms \cite{Le.al.12,Le.15}. First laws of binary mechanics have also
been established in the framework of black hole perturbation theory and
the GSF, first in the case of corotating binaries \cite{GrLe.13}, later
for a test mass on generic bound orbits around a Kerr black hole \cite{Le.14},
and more recently for a massive point particle in Kerr spacetime,
including all conservative GSF effects \cite{Fu.al.17}.

The first law of compact binary mechanics involves the so-called
``redshift'' factor of each point particle, first introduced in
Refs.~\cite{De.08,Sa.al.08,Bl.al.10,Bl.al2.10} for circular orbits,
and later generalized to eccentric orbits
\cite{BaSa.11,Ak.al.15}. Remarkably, the law can be used to relate the
redshift of one of the bodies to the binary's binding energy and
angular momentum, as well as to the relativistic periastron advance
for circular orbits. Since GSF calculations can now compute the
redshift, either numerically with high accuracy
\cite{Sh.al.11,Sh.al.12,Sh.al.14,vdMSh.15}, or analytically to high PN
orders
\cite{BiDa.13,BiDa.14,BiDa2.14,BiDa.15,Jo.al.15,Ka.al.15,Ho.al.16,Ka.al.16},
this translates into new information about the binary's binding energy
and angular momentum, and about the circular-orbit periastron advance,
which can also be computed directly within the GSF framework
\cite{Ba.al.10,Le.al.11,vdM.17}. Summarizing, thanks to the latter
properties, the first laws of
Refs.~\cite{Le.al.12,Bl.al.13,Le.15,Fu.al.17} have already been
applied to:
\begin{itemize}
\setlength\itemsep{0.1em}
\item Determine the numerical values of the ``ambiguity parameters'' that
  appeared in the derivations of the 4PN two-body equations of motion
  \cite{JaSc.12,JaSc.13,JaSc.15,Da.al.14,Da.al.16,Be.al.16,Be.al.17};
  \vspace{0.1cm}
\item Calculate the exact GSF contributions to the binding energy and
  angular momentum for circular orbits, thus allowing a
  coordinate-invariant comparison to NR results \cite{Le.al2.12};
\item Compute the shift in frequencies of Schwarzschild and Kerr
  innermost stable circular orbits induced by the conservative part of
  the GSF \cite{BaSa.09,Da.10,BaSa.10,Le.al2.12,Ak.al.12,Is.al.14,vdM.17};
\item Test the cosmic censorship conjecture in a particular scenario
  where a massive particle subject to the GSF falls into a
  Schwarzschild black hole along unbound orbits \cite{CoBa.15,Co.al.15};
\item Calibrate the effective potentials that enter the EOB model for
  circular orbits \cite{Ba.al.12,Ak.al.12} and mildly eccentric
  orbits \cite{AkvdM.16,Bi.al.16,Bi.al2.16}, and spin-orbit couplings
  for spinning binaries \cite{Bi.al.15};
\item Define the analogue of the redshift of a particle for black
  holes in NR simulations, thus allowing further comparisons to PN and
  GSF calculations \cite{Zi.al.16}.
\end{itemize}

Given the relevance of the first laws to explore the dynamics of
binary systems of compact objects, it is important to address the
following question: do these relations still hold when non-local
effects are accounted for, \textit{i.e.}, when the two-body
Hamiltonian becomes a \textit{functional} (and not merely a function)
of the canonical variables? In the present paper, we extend the
derivation of the first law of Ref.~\cite{Le.15} to 4PN order, for
non-spinning binaries, by taking into account the non-locality of the
action due to the tail effect \cite{Da.al.14,Be.al.16}. In particular,
we shall prove that the first law still holds and takes the standard
form, Eq.~\eqref{firstlaw} below, but with a radial action integral
that gets corrected by 4PN terms related to gravitational-wave tails,
as given in Eq.~\eqref{calR}.

As an application of the first law, we derive the periastron advance
for a slightly non-circular orbit, in the limit where the eccentricity
goes to zero, as a function of the averaged redshift, at 4PN order
and to linear order in the mass ratio. Indeed,
Ref.~\cite{Le.15} showed earlier how the first law can be used to
relate the EOB potentials to the averaged redshift for slightly
eccentric orbits. Since the periastron advance for circular orbits is
related to a linear combination of two of these EOB potentials
\cite{Da.10}, this suggests that the eccentric-orbit first law can be
used to relate \textit{directly} the periastron advance to the
averaged redshift, in the limit of a circular orbit. In this paper we
establish such a relation, Eq.~\eqref{enfin} below, by using our
first law valid for the non-local 4PN dynamics, and check that it is
indeed fully consistent with all known results at 4PN order.

The remainder of this paper is organized as follows. In
Sec.~\ref{sec:nonlocal} we provide a summary of the binary's non-local
dynamics at the 4PN order, as formulated in
Refs.~\cite{Be.al.16,Be.al.17}. The first law with non-local tail
effects is derived in Sec.~\ref{sec:firstlaw}, and a key formula
relating the particles' redshifts to the Hamiltonian is established in
Sec.~\ref{sec:redshift}. Finally, in Sec.~\ref{sec:K-z} we use the
first law to relate the GSF contribution to the periastron advance to
that of the averaged redshift in the circular-orbit limit. Three
appendices give some further technical details. Throughout
  this paper we use geometrized units where $G = c = 1$.

\section{Summary of the 4PN non-local dynamics} 
\label{sec:nonlocal}

In this section and the next one, we employ the canonical Hamiltonian
formalism applied to a binary system of non-spinning point masses
$m_a$, with $a = 1,2$. In an arbitrary frame of reference, the
two-body dynamics is described by canonical variables $\bo{y}_a$ and
$\bo{p}_a$. In the center-of-mass frame, the canonical variables are
the relative position $\bo{x} \equiv \bo{y}_1 - \bo{y}_2$ and linear
momentum $\bo{p} \equiv \bo{p}_1 = -\bo{p}_2$. Furthermore,
introducing polar coordinates in the orbital plane, the conjugate
canonical variables read $(r, \varphi, p_r, p_\varphi)$.

At the 4PN order, the Hamiltonian encoding the dynamics of the binary
system is of the form \cite{Da.al.14,Da.al.16,Be.al.16,Be.al.17}
\beq\label{Hamiltonian}
H = H_0(r, p_r, p_\varphi; m_a) + H_\text{tail}[r, \varphi, p_r,
  p_\varphi; m_a] \,.
\eeq
Here, $H_0$ is a local-in-time piece, the sum of many local (or
``instantaneous'') post-Newtonian terms up to 4PN order. This part
does not depend on the coordinate $\varphi$, so that the conjugate
momentum $p_\varphi$ is conserved for the local dynamics. The tail
term represents a 4PN correction which is a non-local
\textit{functional} of the canonical variables [hence the bracket
  notation used in Eq.~\eqref{Hamiltonian}], given by
\beq\label{Htail} H_\text{tail} = - \frac{M_\text{ADM}}{5}
\,\hat{I}_{ij}^{(3)}\hat{\mathcal{T}}_{ij}^{(3)}\,. \eeq
Because this contribution is a small 4PN correction, thereafter we
will always approximate the ADM mass by the total mass, \textit{i.e.}
$M_\text{ADM} = m \equiv m_1 + m_2$. In the tail term \eqref{Htail},
$\hat{I}_{ij}^{(3)}(t)$ denotes the third time derivative of the
quadrupole moment $I_{ij}(t)$, but with accelerations order reduced by
means of the (Newtonian) equations of motion, which we indicate with
the hat notation. We explicitly have
\beq\label{Iij3} \hat{I}_{ij}^{(3)}(t) = - \frac{2m}{r^2}\left( p_r \,
n^{\langle i}n^{j \rangle} + \frac{4p_\varphi}{r} \, n^{\langle i}\lambda^{j \rangle}\right) ,
\eeq
where the two unit vectors that span the orbital plane are $\bo{n}
\equiv \bo{x}/r = (\cos\varphi, \sin\varphi, 0)$ and $\bm{\lambda}
\equiv (-\sin\varphi, \cos\varphi, 0)$, the angular brackets
surrounding indices denoting the symmetric and trace-free (STF)
projection. The non-local tail factor in Eq.~\eqref{Htail} is given by
\cite{Be.al.16,Be.al.17}
\beq\label{Tijs} \hat{\mathcal{T}}_{ij}^{(s)}(t) =
\mathop{\text{Pf}}_{2r} \int_{-\infty}^{+\infty} \frac{\ud t'}{\vert
  t-t'\vert} \, \hat{I}_{ij}^{(s)}(t') \,, \eeq
with $s=3$ or $4$ in this paper. It involves Hadamard's partie finie
prescription (denoted Pf), which depends on some cut-off scale,
chosen to be the coordinate separation at the current time,
$r=r(t)$. More explicitly, we have
\beq\label{Tijs2} \hat{\mathcal{T}}_{ij}^{(s)}(t) = - 2
\hat{I}_{ij}^{(s)}(t) \ln{r(t)} + \int_0^{+\infty} \! \ud\tau \,
\ln{\left( \frac{\tau}{2} \right)} \left[ \hat{I}_{ij}^{(s+1)}(t-\tau)
  - \hat{I}_{ij}^{(s+1)}(t+\tau) \right] .  \eeq
The tail term \eqref{Htail} depends on the orbital phase $\varphi$, so
that $p_\varphi$ is no longer conserved for the non-local
dynamics. The dependence on the masses $m_a$ is explicit through
Eq.~\eqref{Iij3} and the ADM mass, which reduces to $m=m_1+m_2$ at
this order of approximation.

The non-local in time dynamics of the binary system of point masses
follows from varying the non-local action
\beq\label{S}
S = \int \ud t \, \bigl[ \dot{r} \, p_r + \dot\varphi \, p_\varphi - H
  \bigr] \, ,
\eeq
where the overdot stands for the derivative with respect to the
coordinate time $t$. This yields ordinary looking Hamiltonian
equations,
\beq\label{Heqs}
	\dot{r} = \frac{\delta H}{\delta p_r}\,,\qquad
	\dot{\varphi} = \frac{\delta H}{\delta p_\varphi}\,,\qquad
	\dot{p}_r = - \frac{\delta H}{\delta r}\,,\qquad
	\dot{p}_\varphi = - \frac{\delta H}{\delta \varphi}\,,
\eeq
except that the partial derivatives of the Hamiltonian with respect to
the canonical variables are properly replaced by \textit{functional}
derivatives, in order to account for the non-locality. The functional
derivative of the tail term \eqref{Htail} with respect to $r$ reads as
\beq\label{dHtaildr}
\frac{\delta H_\text{tail}}{\delta r} = - \frac{2m}{5}
  \biggl[\frac{\partial \hat{I}_{ij}^{(3)}}{\partial
      r}\hat{\mathcal{T}}_{ij}^{(3)} - \frac{1}{r}
    \hat{I}_{ij}^{(3)} \hat{I}_{ij}^{(3)} \biggr] \, .
\eeq
It involves the partial derivative of the third (order reduced) time
derivative of the quadrupole moment \eqref{Iij3}. The second term in
the right-hand side of Eq.~\eqref{dHtaildr} comes from the derivative
acting on the Hadamard partie finie scale $r$. Similarly, for the
other variables we have
\beq\label{dHtaildphi}
\frac{\delta H_\text{tail}}{\delta (\varphi,p_r,p_\varphi)} = -
\frac{2m}{5} \,\frac{\partial \hat{I}_{ij}^{(3)}}{\partial
  (\varphi,p_r,p_\varphi)}\hat{\mathcal{T}}_{ij}^{(3)}\,,
\eeq
while the ``functional'' derivative with respect to the masses
obviously reduces to an ordinary derivative, simply given by
\beq\label{varHtailma}
\frac{\delta H_\text{tail}}{\delta m_a} = \frac{\partial
  H_\text{tail}}{\partial m_a} = -
\frac{3}{5}\hat{I}_{ij}^{(3)}\hat{\mathcal{T}}_{ij}^{(3)}\,.
\eeq

Next, we compute the time derivative of the non-local Hamiltonian
\eqref{Hamiltonian} ``on-shell,'' \textit{i.e.} when the field equations
\eqref{Heqs} are satisfied, and obtain \cite{Be.al.17}
\beq\label{dotH}
	\dot{H} = \frac{m}{5}
        \left[\hat{I}_{ij}^{(4)}\hat{\mathcal{T}}_{ij}^{(3)} -
          \hat{I}_{ij}^{(3)}\hat{\mathcal{T}}_{ij}^{(4)}\right] .
\eeq
Hence, for the dynamics deriving from the non-local Hamiltonian
\eqref{Hamiltonian}, the conserved energy $E$, such that $\ud E/\ud
t=0$, differs from the on-shell value of $H$, and we have instead
\cite{Be.al.17}
\beq\label{E}
E=H+\Delta H^\text{DC} + \Delta H^\text{AC}\,,
\eeq
where the first correction is a constant (DC) contribution, while the
second correction is an oscillatory (AC) contribution. The constant
piece turns out to be proportional to the total averaged
gravitational-wave energy flux $\mathscr{F}$,
\beq\label{HDC}
\Delta H^\text{DC} = -
\frac{2m}{5} \, \big\langle \hat{I}_{ij}^{(3)} \hat{I}_{ij}^{(3)} \big\rangle
= - 2 m\,\mathscr{F} \,.
\eeq
The AC piece, on the other hand, is defined to have zero average,
$\langle\Delta H^\text{AC}\rangle=0$, and it must necessarily satisfy
$\ud (\Delta H^\text{AC}) / \ud t = - Q_H$, where $Q_H$ denotes the
right-hand side of Eq.~\eqref{dotH}. From these two requirements, it
follows that
\beq\label{HAC}
\Delta H^\text{AC}(t) = \Big\langle \int_t^u \ud s \,Q_H (s) \,
\Big\rangle_u \,,
\eeq
where ${\langle\rangle}_u$ denotes the average with respect to the
variable $u$, as defined by Eq.~\eqref{longtermaverage} below. In Ref.~\cite{Be.al.17},
an explicit expression for the AC term is given by means of a discrete
Fourier series, using the known Fourier coefficients of the quadrupole
moment as a function of the orbit's eccentricity $e$ (to Newtonian
order). The Fourier series of the AC term is also provided in
Eq.~\eqref{HAC_Fourier} of App.~\ref{appA} below, together with
further details.

Similar results hold for the angular momentum. The Hamilton equation
for $p_\varphi$ reads
\beq\label{dotpphi} \dot{p}_\varphi = \frac{2m}{5} \,\frac{\partial
  \hat{I}_{ij}^{(3)}}{\partial \varphi}\hat{\mathcal{T}}_{ij}^{(3)}\,,
\eeq
showing that $p_\varphi$ is not conserved because of the non-local
tail term. The conserved angular momentum $L$, such that $\ud L/\ud
t=0$, is then obtained in the form
\beq\label{L}
L=p_\varphi+\Delta p_\varphi^\text{DC}+\Delta p_\varphi^\text{AC}\,.
\eeq
The constant DC part is related to the averaged gravitational-wave
flux of angular momentum, $\mathscr{G}$, while the
oscillating AC part is determined by the requirements that
$\langle\Delta p_\varphi^\text{AC}\rangle=0$, and that it must satisfy
$\ud (\Delta p_\varphi^\text{AC}) / \ud t = - Q_{p_\varphi}$, where
$Q_{p_\varphi}$ is the right-hand side of Eq.~\eqref{dotpphi}. More
explicitly, we have
\begin{subequations}
\begin{align}
\Delta p_\varphi^\text{DC} &=
\frac{2m}{5}\,\Big\langle\,\frac{\partial \hat{I}_{ij}^{(3)}}{\partial
  \varphi}\hat{I}_{ij}^{(2)}\,\Big\rangle = - 2
m\,\mathscr{G}\,, \label{pphiDC} \\ \Delta
p_\varphi^\text{AC}(t) &= \Big\langle \int_t^u \ud s
\,Q_{p_\varphi}(s) \,\Big\rangle_u \,. \label{pphiAC}
	\end{align}
\end{subequations}
See App.~\ref{appA} for the Fourier decomposition of $\Delta
p_\varphi^\text{AC}$.

\section{Derivation of the first law} 
\label{sec:firstlaw}

In this section, starting from the non-local Hamiltonian
\eqref{Hamiltonian}, we shall derive a first law of compact binary
mechanics that accounts for the effects of the non-local tail term
\eqref{Htail} at 4PN order. To do so, we start by considering the
unconstrained variation of the Hamiltonian \eqref{Hamiltonian} induced
by infinitesimal changes $\delta r$, $\delta \varphi$, $\delta p_r$,
$\delta p_\varphi$ and $\delta m_a$ of the canonical variables and
component masses, namely
\beq\label{deltaH0}
	\delta H = \frac{\partial H_0}{\partial r} \, \delta r +
        \frac{\partial H_0}{\partial p_r} \, \delta p_r +
        \frac{\partial H_0}{\partial p_\varphi} \, \delta p_\varphi +
        \sum_a \frac{\partial H_0}{\partial m_a} \, \delta m_a +
        \delta H_\text{tail}\,.
\eeq
Here, we separated out the variation of the local instantaneous piece
$H_0$ from that of the non-local tail part. Next, we consider the case
where the changes $(\delta H,\delta r, \delta \varphi, \delta p_r,
\delta p_\varphi, \delta m_a,\delta H_\text{tail})$ correspond to two
neighbouring solutions of the binary's Hamiltonian dynamics. In this
case, one must be careful to perform the variation of the tail
  term \eqref{Htail} on-shell, \textit{i.e.} after having replaced
  into it the motion by a solution of the Hamiltonian equations
  \eqref{Heqs}. That variation is then given by
\beq\label{deltaHtail} \delta H_\text{tail} = - \frac{m}{5}
\left[\delta\hat{I}_{ij}^{(3)}\hat{\mathcal{T}}_{ij}^{(3)} +
  \hat{I}_{ij}^{(3)}\delta\hat{\mathcal{T}}_{ij}^{(3)} + \frac{\delta
    m}{m} \hat{I}_{ij}^{(3)} \hat{\mathcal{T}}_{ij}^{(3)} \right] ,
\eeq
where $\delta\hat{I}_{ij}^{(3)}$ is the variation of the (order
reduced) third \vspace{-0.1cm} time derivative of the quadrupole
moment \eqref{Iij3} with respect to the independent variables and
masses, while $\delta\hat{\mathcal{T}}_{ij}^{(3)}$ denotes the
variation of the onshell value of the tail factor
\eqref{Tijs}--\eqref{Tijs2}. While comparing two neighbouring
solutions of the dynamics, we can also substitute Hamilton's equations
\eqref{Heqs} into Eqs.~\eqref{deltaH0}--\eqref{deltaHtail}, together
with the explicit expressions \eqref{dHtaildr}--\eqref{varHtailma} for
the tail term. A straightforward calculation then yields
\beq\label{deltaH} \delta H = \dot{\varphi}\,\delta p_\varphi -
\dot{p}_\varphi\,\delta \varphi + \dot{r}\,\delta p_r -
\dot{p}_r\,\delta r + \sum_a z_a\,\delta m_a - \frac{2m}{5}
\hat{I}_{ij}^{(3)} \hat{I}_{ij}^{(3)} \, \frac{\delta r}{r} +
\frac{m}{5} \left[\delta\hat{I}_{ij}^{(3)}\hat{\mathcal{T}}_{ij}^{(3)}
  - \hat{I}_{ij}^{(3)}\delta\hat{\mathcal{T}}_{ij}^{(3)}\right] .
\eeq
Notice the last term in square brackets, which is similar to the first
two terms in the right-hand side of Eq.~\eqref{deltaHtail}, but with a
crucial minus sign difference. Finally, in Eq.~\eqref{deltaH} we have
defined the ``redshift'' factor $z_a$ to be the derivative of the
Hamiltonian with respect to the mass $m_a$, namely
\beq\label{zadef} 
z_a \equiv \frac{\partial H}{\partial m_a} = \frac{\partial
  H_0}{\partial m_a} -
\frac{3}{5}\hat{I}_{ij}^{(3)}\hat{\mathcal{T}}_{ij}^{(3)}\,,
\eeq
where we used Eq.~\eqref{varHtailma}. The fact that the quantity
\eqref{zadef} is indeed the redshift factor of the particle $a$,
namely that $z_a=\ud\tau_a/\ud t$, is not trivial and will be proven
in Sec.~\ref{sec:redshift} below.

Next, to simplify the tail terms in square brackets in the right-hand
side of Eq.~\eqref{deltaH}, we make use of the explicit Fourier series
representations of the quadrupole moment and of the tail factor
\eqref{Tijs}, which are given by the formulas \eqref{Iij_Fourier} and
\eqref{tailfactorFourier} in App.~\ref{appA}. Of course, this is
allowed since we are considering the on-shell variation of the tail
term. It can then easily be shown that
\beq\label{diff}
  \delta\hat{I}_{ij}^{(3)}\hat{\mathcal{T}}_{ij}^{(3)} -
  \hat{I}_{ij}^{(3)}\delta\hat{\mathcal{T}}_{ij}^{(3)} = 2
  \hat{I}_{ij}^{(3)} \hat{I}_{ij}^{(3)} \! \left( \frac{\delta r}{r} +
  \frac{\delta n}{n} \right) + \Delta \,, \eeq
where $n \equiv 2\pi / P$ denotes the frequency associated with
the period $P$ of the radial motion, while the extra piece $\Delta$
represents a more complicated expression, involving a double Fourier
series over the Fourier components of the quadrupole moment and
their variations,
\beq\label{Delta} \Delta = 2 \sum_{p,q}
\mathop{{\mathcal{I}}}_{p}{}_{\!\!ij} n^3 \ue^{\ui p
  \ell}\,\delta\Bigl(\mathop{{\mathcal{I}}}_{q}{}_{\!\!ij} n^3
\ue^{\ui q \ell}\Bigr) \, p^3 q^3 \ln\left|\frac{p}{q}\right| \,. \eeq
To be clear, we are considering the difference between two
infinitesimally close configurations associated with quadrupole
moments $I_{ij}(t)$ and $I'_{ij}(t)$. These configurations have
different radial frequencies $n$ and $n'$, semi-major axes $a$ and
$a'$, and eccentricities $e$ and $e'$, as well as different
masses. The Fourier decomposition \eqref{Delta} involves the Fourier
coefficients ${}_p\mathcal{I}_{ij}$ and ${}_p\mathcal{I}'_{ij}$ (see
App.~\ref{appA} for definitions) and different mean anomalies
$\ell=n(t-t_0)$ and $\ell'=n'(t-t'_0)$. We denote
$\delta{}_p\mathcal{I}_{ij}={}_p\mathcal{I}'_{ij}-{}_p\mathcal{I}_{ij}$,
$\delta n = n'-n$, and so on, \textit{e.g.}, $\delta\ue^{\ui p \ell} =
\ue^{\ui p \ell'}-\ue^{\ui p \ell}$. Since the tail factor
\eqref{Tijs} occurs at 4PN order, we can compute these configurations
using Newtonian elliptical orbits.

Following Ref.~\cite{Le.15} we shall now consider the time average of
the variational formula \eqref{deltaH}. In the most general case, the
time average $\langle f \rangle$ of a given function $f(t)$ will be
defined as
\beq\label{longtermaverage} \langle f \rangle \equiv \lim_{T\to
  +\infty}\frac{1}{2T}\int_{-T}^{T}\ud t\,f(t) \, .  \eeq
But for periodic functions with period $P$, this reduces to the usual
average $\langle f \rangle = \frac{1}{P} \int_0^P\ud t\,f(t)$ over one
radial period. Let us first check that the time average of the
quantity \eqref{Delta} is zero. Indeed, expanding the variational
$\delta$ operation, it is clear that all the terms proportional to
$\ue^{\ui (p+q) \ell}$ average to zero, $\langle\ue^{\ui (p+q)
  \ell}\rangle=0$, since $p+q\not= 0$ thanks to the presence of the
logarithmic factor $\ln\vert p/q\vert$. But we also have terms
proportional to $\ue^{\ui p \ell}\delta\ue^{\ui q \ell} = \ue^{\ui (p
  \ell + q \ell')}-\ue^{\ui (p+q) \ell}$. However, recall that the two
configurations we consider are infinitesimally close, so we have $p n
+ q n'\not= 0$ in this case. Then, by applying the long-time average
\eqref{longtermaverage} we readily obtain $\langle\ue^{\ui (p \ell + q
  \ell')}\rangle=0$. Finally, we conclude that the quantity
\eqref{Delta} has, indeed, zero average:
$\langle\Delta\rangle=0$. Therefore, substituting Eq.~\eqref{diff}
into the variational formula \eqref{deltaH} and averaging, we obtain
\beq\label{deltaE0} \langle \delta H \rangle = \langle
\dot{\varphi}\,\delta p_\varphi - \dot{p}_\varphi\,\delta
\varphi\rangle + \langle\dot{r}\,\delta p_r - \dot{p}_r\,\delta
r\rangle + \sum_a \langle z_a\rangle\,\delta m_a + 2m \frac{\delta
  n}{n}\mathscr{F} \,, \eeq
where we used the fact that $n$ and $m_a$ are constant, while the last
term contains the averaged gravitational-wave flux of energy
$\mathscr{F} = \frac{1}{5}\av{\hat{I}_{ij}^{(3)} \hat{I}_{ij}^{(3)}}$.

To evaluate the radial contribution, we proceed as in
Ref.~\cite{Le.15}. Since the average of the time derivative of a
periodic function vanishes, the radial contribution to
Eq.~\eqref{deltaE0} can be written as
\beq\label{ploup}
\langle\dot{r} \, \delta p_r - \dot{p}_r \, \delta r\rangle =
\langle\delta(\dot{r}p_r)\rangle = \frac{1}{P}\int_{0}^{P} \! \ud t \,
\delta(\dot{r}p_r) = \frac{2}{P}\int_{r_-}^{r_+} \! \delta(p_r\ud r) \, ,
\eeq
where $r_-$ and $r_+$ denote the orbit's periastron and apastron, at
which $\dot{r} = 0$. Next, we can pull out the variation $\delta$ from
the integral. To see this, it is convenient to write Eq.~\eqref{ploup}
as an integral over the complex plane, initially along the segment
$[r_-, r_+]$ along the real axis, but then deformed into an integral
over a given closed contour $C$ surrounding $r_-$ and $r_+$ in the
complex plane, say $\frac{1}{P}\oint_C\delta(p_r\ud r)$. When doing
so, since the contour is fixed, one can ignore the variation of $r_-$
and $r_+$ in the process. This is Sommerfeld's well known method of
contour integrals; see \textit{e.g.} Ref.~\cite{DaSc.88} or App.~C in
\cite{Be.al.17}. Finally, we get
\beq\label{contrib1} \langle\dot{r} \, \delta p_r - \dot{p}_r \,
\delta r\rangle = \frac{1}{P} \, \delta \! \oint_C p_r \, \ud r =
n\,\delta R \, , \eeq
where we recall that $n=2\pi/P$ is the radial frequency, or mean
motion, and where $R$ is the radial action integral, defined by
\beq\label{radialR} R \equiv \frac{1}{2\pi} \oint p_r\,\ud r =
\frac{1}{\pi} \int_{r_-}^{r_+} \! p_r\,\ud r \,.  \eeq

Now, to evaluate the azimuthal contribution to \eqref{deltaE0}, we
recall that $p_\varphi$ is not conserved in the non-local case [see
  Eq.~\eqref{dotpphi}], such that the result of the calculation will
not reduce to the usual $\omega \, \delta L$ term. Instead, we write
\begin{align}\label{contrib2}
\langle \dot{\varphi} \, \delta p_\varphi - \dot{p}_\varphi \, \delta
\varphi\rangle &= \av{\dot{\varphi} \, \delta L} - \big\langle
\dot{\varphi} \, \delta \Delta p_\varphi^\text{DC} \big\rangle -
\big\langle \dot{\varphi} \, \delta \Delta p_\varphi^\text{AC}
\big\rangle + \big\langle \bigl(\Delta p_\varphi^\text{AC}
\bigr)\dot{} \, \delta \varphi \big\rangle \nonumber \\ &= \omega \,
\delta L - \omega \, \delta \Delta p_\varphi^\text{DC} - \langle
\delta(\dot{\varphi}\Delta p_\varphi^\text{AC}) \rangle \, ,
\end{align}
in which we used Eq.~\eqref{L} as well as the fact that $L$ and
$\Delta p_\varphi^\text{DC}$ are both constant, and we introduced the
orbital-averaged azimuthal frequency
\beq\label{omega} \omega \equiv \langle \dot{\varphi} \rangle =
\frac{1}{P} \int_0^P \! \ud t \, \dot{\varphi} = n \oint
\frac{\ud\varphi}{2\pi} = K n \,, \eeq
where $2\pi K \equiv \oint \ud\varphi = 2\pi + \Delta \Phi$ is the
accumulated azimuthal angle per radial period, with $\Delta\Phi$ the
relativistic periastron advance. In the second line of
Eq.~\eqref{contrib2}, we may then use \eqref{pphiDC} in the second
term and handle the last term just like the radial contribution
\eqref{contrib1}, such that finally
\beq\label{contrib2bis} \langle \dot{\varphi} \, \delta p_\varphi -
\dot{p}_\varphi \, \delta \varphi\rangle = \omega \, \delta L + \omega
\, \delta \bigl( 2m \, \mathscr{G} \bigr) - n \, \delta \biggl(
\frac{1}{2\pi} \! \oint \Delta p_\varphi^\text{AC}\,\ud \varphi
\biggr) \, .  \eeq

At last, we have to take into account the relationship \eqref{E}, which
implies that the term $\av{\delta H}$ in Eq.~\eqref{deltaE0} is not simply
equal to $\delta E$. Instead, the conserved energy $E$ (which includes
the total rest mass $m = m_1 + m_2$) gets shifted by the DC correction
\eqref{HDC}, while the AC correction \eqref{HAC} does not contribute
since it has zero time average:
\beq\label{contrib3} \av{\delta H} = \delta \av{H} = \delta E + \delta
\bigl( 2m \, \mathscr{F} \bigr) \, .  \eeq
Finally, collecting the intermediate results \eqref{deltaE0},
\eqref{contrib1}, \eqref{contrib2bis} and \eqref{contrib3}, and
combining the 4PN contributions that involve the gravitational-waves
fluxes $\mathscr{F}$ and $\mathscr{G}$, where at that order of
approximation one may replace $n$ by $\omega$ and $\delta n$ by
$\delta \omega$ if needed, we obtain a first law of binary
mechanics that takes the standard form, as established in
Ref.~\cite{Le.15}, namely
\beq\label{firstlaw} \delta E = \omega \,\delta L + n \,\delta
\mathscr{R} + \sum_a \langle z_a\rangle\,\delta m_a \,, \eeq
but where, as anticipated above, the radial action integral
\eqref{radialR} gets corrected at 4PN order by terms originating from
the non-local tail:
\beq\label{calR} \mathscr{R} = R + 2m \biggl( \mathscr{G} -
  \frac{\mathscr{F}}{\omega} \biggr) - \frac{1}{2\pi} \oint \Delta
p_\varphi^\text{AC}\,\ud \varphi \,. \eeq
Heuristically, one may interpret the additional contributions
proportional to the gravitational-wave fluxes as being related to the
energy and angular momentum content in gravitational waves in the far
zone.\footnote{Note that the gravitational-wave fluxes are themselves
  related by a first law in the adiabatic approximation, namely
  $\mathscr{F} = \omega \, \mathscr{G} - n \, \langle \dot{R}\rangle -
  \sum_a \varepsilon_a \langle z_a\rangle \mathscr{H}_a$; see Sec.~V A
  in Ref.~\cite{Le.15}.} Moreover, we recall that $2m \mathscr{F} = -
\Delta H^\text{DC}$ and $2m \mathscr{G} = - \Delta
p_\varphi^\text{DC}$. The Fourier decomposition of the last term in
the right-hand side of Eq.~\eqref{calR} is investigated in
App.~\ref{appA}. Importantly, we note that the correction terms in
\eqref{calR} vanish for circular orbits, because for such orbits the
Newtonian gravitational-wave fluxes obey $\mathscr{F} = \omega \,
\mathscr{G}$, while $\dot{\varphi}$ is constant and $\big\langle
\Delta p_\varphi^\text{AC} \big\rangle = 0$. Hence, the circular-orbit
condition $R = 0$ implies $\mathscr{R} = 0$.

The authors of Refs.~\cite{Da.al.15,Da.al.16} discussed how the
non-local Hamiltonian \eqref{Hamiltonian}--\eqref{Htail} can formally
be reduced to an ordinary \textit{local} Hamiltonian by means of a
suitable transformation $(r,\varphi,p_r,p_\varphi) \longrightarrow
(r^\text{loc},\varphi^\text{loc},p^\text{loc}_r,p^\text{loc}_\varphi)$
of the phase-space variables. Having performed such a ``localization''
of the Hamiltonian, one could then follow Ref.~\cite{Le.15} to derive
an ordinary first law of binary mechanics. That ``local'' law would be
identical to our Eq.~\eqref{firstlaw}, except that the radial action
integral therein, say $R^\text{loc}$, would be given by the usual
expression defined in terms of the shifted variable
$p^\text{loc}_r$. Of course, our modified radial action integral
$\mathscr{R}$ obtained in Eq.~\eqref{calR} should be identical to the
local radial action integral $R^\text{loc}$ when it is expressed in
terms of the natural invariants $E$ and $L$ (and masses $m_a$), namely
\beq\label{locnonlocal} \mathscr{R}(E,L) = R^\text{loc}(E,L) \equiv
\frac{1}{2\pi} \oint \ud r^\text{loc}\,p^\text{loc}_r(r^\text{loc}, E,
L) \,.  \eeq

Before closing this section, we note that one can easily derive a
``first integral'' relationship associated with the variational first
law \eqref{firstlaw}, namely
\beq\label{firstintegral} E = 2 \omega L + 2 n \mathscr{R} + \sum_a
m_a \langle z_a\rangle \,.  \eeq
This can be proven in various ways. For instance, one might notice
that $E$ is an homogeneous function of degree one in the variables
$\sqrt{L}$, $\sqrt{\mathscr{R}}$ and $m_a$, such that
\eqref{firstintegral} comes from applying Euler's theorem for
homogeneous functions; see Refs.~\cite{Le.al.12,Bl.al.13,Le.15}.

\section{Derivation of the redshift factor} 
\label{sec:redshift}

In this section we shall prove that the quantity $z_a$ defined by
Eq.~\eqref{zadef} actually coincides with the redshift $\ud \tau_a /
\ud t$ of the particle $a$. Our proof will be based on the use of the
Fokker Lagrangian, and is a minor adaptation of the proof already
given in Ref.~\cite{Bl.al.13}, with the simplification that we
consider here only non-spinning particles, but with the slight
complication that the dynamics is non-local because of the 4PN tail
effect.

The Fokker Lagrangian of a system of point particles was defined,
\textit{e.g.}, in Ref.~\cite{Be.al.16}. We start from the
gravitation-plus-matter Lagrangian of general relativity,
\beq\label{Lag} L = L_\text{g}\bigl[g_{\mu\nu}\bigr] +
L_\text{m}\bigl[g_{\mu\nu};\bo{y}_a,\bo{v}_a;m_a\bigr]\,.  \eeq
The gravitational part $L_\text{g}$ is the usual Einstein-Hilbert term,
written in the Landau-Lifshitz form, with the harmonic
gauge-fixing term; see Eq.~(2.1) in Ref.~\cite{Be.al.16}. The matter
Lagrangian for the system of point particles is given by
\beq\label{Lm} L_\text{m}\bigl[g_{\mu\nu};\bo{y}_a,\bo{v}_a;m_a\bigr]
= - \sum_a m_a \sqrt{-g_{\mu\nu}(y_a) v_a^\mu v_a^\nu}\,, \eeq
where $y_a^\mu=(t,\bo{y}_a)$ and $v_a^\mu=(1,\bo{v}_a)$ denote the
trajectories and ordinary coordinate velocities, with $\bo{v}_a(t)
\equiv \dot{\bo{y}}_a(t)$, and $g_{\mu\nu}(y_a)$ stands for the metric
evaluated at the location of the particle $a$, following some
regularization scheme, in principle dimensional
regularization \cite{Be.al.16}.

The Einstein field equations in harmonic coordinates follow from
varying the Lagrangian \eqref{Lag} with respect to the metric. These
equations are then solved perturbatively, yielding an explicit
PN-iterated harmonic-coordinates solution, say
\beq\label{gbar}
\overline{g}_{\mu\nu}(x) \equiv
\overline{g}_{\mu\nu}(\bo{x};\bo{y}_b,\bo{v}_b,\bo{a}_b;m_b)\,.
\eeq
This solution depends on the positions $\bo{y}_b$ and velocities
$\bo{v}_b$ of all of the particles, but also on their accelerations
and any possible derivatives of accelerations that can get generated
at high PN orders, and are here symbolized by
$\bo{a}_b\equiv(\dot{\bo{v}}_b,\ddot{\bo{v}}_b,\cdots)$. Of course,
the solution \eqref{gbar} depends also on all the masses $m_b$. The
Fokker Lagrangian is then defined by inserting the explicit PN
solution \eqref{gbar} back into the Lagrangian \eqref{Lag}, thus
obtaining
\beq\label{Fokker}
L_\text{F}\bigl[\bo{y}_a,\bo{v}_a,\bo{a}_a;m_a\bigr] \equiv
L_\text{g}\bigl[\overline{g}_{\mu\nu}
  (\bo{x};\bo{y}_b,\bo{v}_b,\bo{a}_b;m_b)\bigr] +
L_\text{m}\bigl[\overline{g}_{\mu\nu}
  (\bo{y}_a;\bo{y}_b,\bo{v}_b,\bo{a}_b;m_b);\bo{v}_a,m_a\bigr]\,.
\eeq
This Lagrangian is a generalized Lagrangian, depending not only on
positions and velocities, but also on accelerations and derivatives of
accelerations. Taking the functional derivative with respect to the
position of the particle $a$ yields
\beq
\frac{\delta L_\text{F}}{\delta \bo{y}_a} = \frac{\delta
  \overline{g}_{\mu\nu}}{\delta \bo{y}_a} \frac{\delta L}{\delta
  g_{\mu\nu}}{\bigg|}_{\overline{g}_{\mu\nu}} + \frac{\delta
  L_\text{m}}{\delta \bo{y}_a}{\bigg|}_{\overline{g}_{\mu\nu}}\,.
\eeq
But since $\delta L/\delta g_{\mu\nu}=0$ holds for the actual PN
solution $\overline{g}_{\mu\nu}$ of the Einstein field equations, the
basic property of the Fokker Lagrangian follows, namely that its
functional derivative with respect to one of the particle's position
reduces to that of the matter Lagrangian while holding the metric
fixed in Eq.~\eqref{Lm}:
\beq\label{dLdy}
\frac{\delta L_\text{F}}{\delta \bo{y}_a} =\frac{\delta
  L_\text{m}}{\delta \bo{y}_a}{\bigg|}_{\overline{g}_{\mu\nu}}\,.
\eeq
Therefore, $\delta L_\text{F}/\delta \bo{y}_a = 0$ yields the correct
equations of motion for the system of point masses in the metric
generated by the particles themselves. 

Next, we can apply the very same argument for the variation of the
Fokker Lagrangian with respect to the mass $m_a$, holding $\bo{y}_b$,
$\bo{v}_b$, $\bo{a}_b$ fixed. We find that the dependence over the
mass that is hidden into the PN solution $\overline{g}_{\mu\nu}$ gets
cancelled by the fact that $\delta L/\delta
g_{\mu\nu}\vert_{\overline{g}_{\mu\nu}} = 0$. Hence we obtain the
important result
\beq\label{dLdm}
\frac{\delta L_\text{F}}{\delta m_a} = \frac{\delta L_\text{m}}{\delta
  m_a}{\bigg|}_{\overline{g}_{\mu\nu}} \,.
\eeq
As is clear from Eq.~\eqref{Lm}, the functional derivative of the
matter Lagrangian at fixed $\overline{g}_{\mu\nu}$ in the right-hand
side of \eqref{dLdm} reduces to an ordinary derivative, and we get
\beq\label{dLdmexpl}
\frac{\delta L_\text{F}}{\delta m_a} = - \sqrt{-g_{\mu\nu}(y_a)
  v_a^\mu v_a^\nu}\,.
\eeq

Finally, it remains to go from the Fokker Lagrangian $L_\text{F}$ to
the corresponding Hamiltonian $H_\text{F}$. The only subtlety is that
the harmonic-coordinates Fokker Lagrangian is a generalized Lagrangian.
Hence we must first get rid of the accelerations by performing suitable
shifts of the trajectories, so as to obtain an ordinary Lagrangian,
depending only on the positions and velocities. Such shifts have
recently been performed in Ref.~\cite{Be.al.16}, and discussed in a
more general context in Ref.~\cite{DaSc.91}; notice that the 4PN tail
term is also transformed into an ordinary---although still
non-local---term by applying suitable shifts. Now, the new metric
expressed in the new, shifted variables will take the same form as in
\eqref{gbar}, but without accelerations, because the redefinition of
the trajectories can be seen as being induced by a coordinate
transformation of the ``bulk'' metric. Hence the derivation given
above applies to the new Lagrangian with shifted variables, and the
relationship \eqref{dLdm} still holds.  Furthermore, that Lagrangian
being ordinary, a usual Legendre transformation can be performed to
define the Hamiltonian as $H_\text{F} \equiv \sum_a p_a^i v_a^i -
L_\text{F}$, where $p_a^i=\delta L_\text{F}/\delta v_a^i$ gives
$\bo{v}_a$ as a functional of the canonical positions $\bo{y}_b$ and
momenta $\bo{p}_b$. From the properties of the Legendre
transformation, we readily find that the derivative of the Hamiltonian
with respect to the mass $m_a$, while holding $\bo{y}_b$ and
$\bo{p}_b$ fixed, is simply
\beq\label{dHdm}
\frac{\delta H_\text{F}}{\delta m_a} = - \frac{\delta
  L_\text{F}}{\delta m_a} = \sqrt{-g_{\mu\nu}(y_a) v_a^\mu v_a^\nu}\,.
\eeq
Here, the velocities are to be considered as functionals of the
canonical variables, $\bo{v}_a[\bo{y}_b,\bo{p}_b]$. Since the Fokker
Hamiltonian $H_\text{F}$ that we have just introduced is precisely the
Hamiltonian \eqref{Hamiltonian} that we considered in
Secs.~\ref{sec:nonlocal} and \ref{sec:firstlaw}, we have proven that
the quantity $z_a$ defined in Eq.~\eqref{zadef} is indeed the redshift
associated with the particle $a$, namely that
\beq\label{redshift}
z_a = \frac{\ud\tau_a}{\ud t} = \sqrt{-g_{\mu\nu}(y_a) v_a^\mu
  v_a^\nu}\,.
\eeq
%

\section{Periastron advance and averaged redshift}
\label{sec:K-z}

Throughout this section we assume that one of the two compact objects,
say body 1, is much less massive than the other, and we work to linear
order in the mass ratio $q \equiv m_1/m_2 \ll 1$, or equivalently to
linear order in the symmetric mass ratio $\nu \equiv m_1 m_2 / m^2 = q
+ \calO(q^2)$. Our objective is to relate, in the circular-orbit limit, the
$\calO(\nu)$ contributions to the periastron advance and to the
averaged redshift $\av{z} \equiv \av{z_1}$ associated with the lighter
body.

A generic bound (eccentric) orbit can be parameterized using the two
orbital frequencies $n$ and $\omega$, or equivalently using $\omega$
and the periastron advance $K = \omega/n$. Hence, to first order in
the symmetric mass ratio $\nu$, we may consider the following
expansions of the modified radial action variable \eqref{calR} and the
averaged redshift of the lighter body:
\begin{subequations}
\begin{align}
\scrR(K,\omega) &= \scrR_{(0)}(K,\omega) + \nu \,
\scrR_{(1)}(K,\omega) + \calO(\nu^2) \, , \label{R}
\\ \av{z}(n,\omega) &= \av{z}_{(0)}(n,\omega) + \nu \,
\av{z}_{(1)}(n,\omega) + \calO(\nu^2) \, ,\label{<z>}
\end{align}
\end{subequations}
where $\scrR_{(0)}$ and $\av{z}_{(0)}$ denote the values of those quantities
in the (Schwarzschild) background, while $\scrR_{(1)}$ and $\av{z}_{(1)}$
represent first-order GSF corrections.

A circular orbit is defined by the condition of a vanishing radial
action: $R = 0$; see \eqref{radialR}. Crucially, as mentionned
earlier, the corrective terms in the right-hand side of \eqref{calR}
vanish in the circular-orbit limit, such that $R = 0$ implies $\scrR =
0$. For the one-parameter family of circular orbits, the frequencies
$n$ and $\omega$ are no longer independant, \textit{i.e.} $n =
n^\text{circ}(\omega)$, or equivalently
\beq\label{Kbar} K = K^\text{circ}(\omega) = K_{(0)}(\omega) + \nu \,
K_{(1)}(\omega) + \calO(\nu^2) \, .  \eeq
Our goal here is to relate the $\calO(\nu)$ contribution to
$K^\text{circ}(\omega)$, namely $K_{(1)}(\omega)$, to the $\calO(\nu)$
contribution $\av{z}_{(1)}(n,\omega) $ to the redshift \eqref{<z>} in the
circular-orbit limit.

Expanding the circular-orbit condition $\scrR=0$ to first order in the
symmetric mass ratio, while using Eqs.~\eqref{R} and~\eqref{Kbar}, we
get
\beq\label{toto} 0 = \scrR_{(0)}\bigl(K_{(0)}(\omega),\omega\bigr) +
\nu \left[ K_{(1)}(\omega) \left( \frac{\partial \scrR_{(0)}}{\partial
    K} \right)_{\!\omega}\!\!\!\bigl(K_{(0)}(\omega),\omega\bigr) +
  \scrR_{(1)}\bigr(K_{(0)}(\omega),\omega\bigr) \right] \! +
\calO(\nu^2) \, .  \eeq
The first term in the right-hand side of~\eqref{toto} vanishes
identically. Because the contribution $\calO(\nu)$ must also vanish
identically, we obtain
\beq\label{pif0} K_{(1)}(\omega) = -
\frac{\scrR_{(1)}\bigl(K_{(0)}(\omega),\omega\bigr)}{\left(
  \frac{\partial \scrR_{(0)}}{\partial K}
  \right)_{\!\omega}\!\!\bigl(K_{(0)}(\omega),\omega\bigr)} \,.  \eeq
At this stage, it gets convenient to treat $K$ as a function of
$\omega$ and $\scrR_{(0)}$, defined by inverting
$\scrR_{(0)}=\scrR_{(0)}(K,\omega)$. Since $\scrR_{(0)}=0$ defines
circular orbits in the background (\textit{i.e.}, when the mass ratio
is $\nu=0$), we can rewrite Eq.~\eqref{pif0} as
\beq\label{pif} K_{(1)}(\omega) = -
\scrR_{(1)}\bigl(K_{(0)}(\omega),\omega\bigr) \left( \frac{\partial
  K}{\partial \scrR_{(0)}}
\right)_{\!\omega}\!\!\bigl(\omega,\scrR_{(0)}=0\bigr) \,.  \eeq
A simple change of variables from $(\omega,\scrR_{(0)})$ to the
frequencies $(\omega,n)$ yields $(\partial K / \partial
\scrR_{(0)})_\omega = (\partial K / \partial n)_\omega (\partial n /
\partial \scrR_{(0)})_\omega = - (K^2 / \omega) (\partial n / \partial
\scrR_{(0)})_\omega$, and here we can replace $K$ by the background
value $K_{(0)}$. Therefore, Eq.~\eqref{pif} can be written in the
convenient form
\beq\label{plouf} K_{(1)}(\omega) = \frac{K^2_{(0)}(\omega)}{\omega}
\left[\scrR_{(1)}\left( \frac{\partial \scrR_{(0)}}{\partial n}
  \right)_{\!\omega}^{-1}\right]
\Bigl(\omega,n\bigl(\omega,\scrR_{(0)}=0\bigr)\Bigr) \, , \eeq
where the right-hand side is computed for $\omega$ and
$n(\omega,\scrR_{(0)}=0)$, which is the radial frequency as a function
of $\omega$ for circular orbits in the Schwarzshild background, say
$n^\text{circ}_{(0)}(\omega)$.

Next, we need to relate $\scrR_{(1)}(\omega,n)$ to the GSF
contribution $\av{z}_{(1)}(n,\omega)$ to the averaged
redshift \eqref{<z>}, in the circular-orbit limit. But from
Eqs.~(5.8b) and (5.9c) of Ref.~\cite{Le.15}, we know that, for any
dimensionless frequencies $(\hat{\omega},\hat{n}) \equiv
(m\omega,mn)$, and up to an irrelevant overall scaling of
$\scrR_{(0)}$ and $\scrR_{(1)}$,
\begin{subequations}\label{babel}
\begin{align}
\scrR_{(0)}(\hat{\omega},\hat{n}) &= - \frac{\partial
  \av{z}_{(0)}}{\partial \hat{n}} \, ,
\\ \scrR_{(1)}(\hat{\omega},\hat{n}) &= - \frac{1}{2} \left(
\frac{\partial \av{z}_{(1)}}{\partial \hat{n}} + \frac{\partial
  \av{z}_{(0)}}{\partial \hat{n}} - \hat{n} \, \frac{\partial^2
  \av{z}_{(0)}}{\partial \hat{n}^2} - \hat{\omega} \, \frac{\partial^2
  \av{z}_{(0)}}{\partial \hat{\omega} \partial \hat{n}} \right) .
	\end{align}
\end{subequations}
These expressions were established from a first law derived starting
from a local Hamiltonian. However, since we proved in
Sec.~\ref{sec:firstlaw} that a similar first law relation holds for
the non-local Hamiltonian \eqref{Hamiltonian}, as long as the radial
action $R$ is replaced by $\scrR$, we conclude that \eqref{babel}
hold when expressed in terms of the corrected radial action $\scrR$
[recall also Eq.~\eqref{locnonlocal}].  Inserting these expressions
into Eq.~\eqref{plouf} yields
\beq\label{K_GSF} K_{(1)}(\hat{\omega}) =
\frac{K^2_{(0)}(\hat{\omega})}{2\hat{\omega}} \left( \frac{\partial^2
  \av{z}_{(0)}}{\partial \hat{n}^2} \right)^{-1} \left( \frac{\partial
  \av{z}_{(1)}}{\partial \hat{n}} + \frac{\partial
  \av{z}_{(0)}}{\partial \hat{n}} - \hat{n} \, \frac{\partial^2
  \av{z}_{(0)}}{\partial \hat{n}^2} - \hat{\omega} \, \frac{\partial^2
  \av{z}_{(0)}}{\partial \hat{\omega} \partial \hat{n}} \right) ,
\eeq
where the right-hand side is still computed at $\omega$ and
$n^\text{circ}_{(0)}(\omega)=n(\omega,\scrR_{(0)}=0)$.

To evaluate more explicitly the latter expression in the
circular-orbit limit, it is especially convenient to parametrize the
orbit in terms of the usual Schwarzschild ``semi-latus rectum'' $p$
and ``eccentricity'' $e$, instead of the frequencies $\hat{n}$ and
$\hat{\omega}$, and to perform a Taylor expansion in the limit where $e \to
0$ (see App.~\ref{appB} for more details). For instance, we write
\beq\label{pouf} \left( \frac{\partial \av{z}_{(1)}}{\partial \hat{n}}
\right)_{\!\hat{\omega}} = \left( \frac{\partial
  \av{z}_{(1)}}{\partial p} \right)_{\!e} \left( \frac{\partial
  p}{\partial \hat{n}} \right)_{\!\hat{\omega}} + \left(
\frac{\partial \av{z}_{(1)}}{\partial e} \right)_{\!p} \left(
\frac{\partial e}{\partial \hat{n}} \right)_{\!\hat{\omega}} \, .
\eeq
Adapting notations, the expressions for $\hat{n}(p,e)$,
$\hat{\omega}(p,e)$ and $\av{z}_{(0)}(p,e)$ are given, for instance,
in Eqs.~(2.4)--(2.10) of Ref.~\cite{Ak.al.15}. These relationships can
be computed analytically, as Taylor expansions in the
eccentricity $e$. We collect all the required results in
App.~\ref{appB}; in particular, the coefficients appearing in
\eqref{pouf} are given in Eq.~\eqref{tac} there.  Moreover in the
small-$e$ limit, the $\calO(\nu)$ contribution to the averaged
redshift can be expanded as\footnote{We employ the Landau symbol $o$
  for remainders with its usual meaning.}
\beq\label{tic} \av{z}_{(1)}(p,e) = z_{(1)}(p) + \frac{e^2}{2} \,
\av{z}^{e^2}_{(1)}(p) + o(e^2) \,, \eeq
where we used the notations \vspace{-0.1cm} $z_{(1)}(p) \equiv
\av{z}_{(1)}(p,0)$ and $\av{z}^{e^2}_{(1)}(p) \equiv (\partial^2
\av{z}_{(1)} / \partial e^2)(p,0)$. Accurate GSF data \vspace{-0.05cm}
for $\av{z}^{e^2}_{(1)}(p)$ were computed for separations $6 < p
\leqslant 1200$ in Refs.~\cite{Ak.al.15,AkvdM.16}. Note that a
contribution linear in the eccentricity cannot appear in
Eq.~\eqref{tic}, otherwise the expression \eqref{K_GSF} for $K_{(1)}$
would be singular in the circular-orbit limit $e \to 0$, as can be
seen from Eqs.~\eqref{pouf} and \eqref{tac2}. Substituting \eqref{tac}
and \eqref{tic} into \eqref{pouf}, we find that both the
circular-orbit contribution $z_{(1)}(p)$ and the leading
finite-eccentricity contribution $\av{z}^{e^2}_{(1)}(p)$ appear in the
final expression for $(\partial \av{z}_{(1)} / \partial
\hat{n})_{\hat{\omega}}$ [and hence will appear in that for
  $K_{(1)}(\hat{\omega})$], namely
\beq\label{bim} \frac{\partial \av{z}_{(1)}}{\partial \hat{n}}
\bigg|_{e = 0} = \frac{4}{3} \, \frac{p^2 \sqrt{p-6}}{4p^2 - 39p +86}
\left[ p \, (p^2 - 10 p + 22) \, \frac{\ud z_{(1)}}{\ud p} -
  (p-2)(p-6) \, \frac{\av{z}^{e^2}_{(1)}(p)}{2} \right] .  \eeq

Finally we need the closed-form expressions of the frequency
derivatives of the background averaged redshift
$\av{z}_{(0)}(\hat{\omega},\hat{n})$ that appear in
Eq.~\eqref{K_GSF}. In the small-$e$ limit, these are given by
Eqs.~\eqref{trucs} in App.~\ref{appB}. Then, combining
Eqs.~\eqref{K_GSF}, \eqref{bim} and \eqref{trucs}, our final
expression for the $\calO(\nu)$ contribution to the periastron advance
simply reads
\beq\label{enfin} K_{(1)}(p) = - \frac{\sqrt{p}}{(p-6)^{3/2}} +
\frac{p\sqrt{p-3}}{(p-6)^{5/2}} \biggl[ p \, (p^2 - 10 p + 22) \,
  \frac{\ud z_{(1)}}{\ud p} - (p-2)(p-6) \,
  \frac{\av{z}^{e^2}_{(1)}(p)}{2} \Biggr] \, .  \eeq
Equivalently, in terms of the quantity $W \equiv 1/ K^2$ that was
introduced in Ref.~\cite{Da.10}, namely $W(x) = 1-6x + \nu \, \rho(x)
+ \mathcal{O}(\nu^2)$, where $x \equiv \hat{\omega}^{2/3} = p^{-1} +
\mathcal{O}(\nu)$, we readily find for the GSF contribution
\beq\label{rho} \rho(x) = 2x + 2 \sqrt{1-3x} \left[
  \frac{1-10x+22x^2}{1-6x} \, \frac{\ud z_{(1)}}{\ud x} + (1-2x) \,
  \frac{\av{z}_{(1)}^{e^2}(x)}{2x} \right] .  \eeq
As an important check of these results, we verified that the formula
\eqref{rho} is recovered when combining the relationship of
Ref.~\cite{Da.10} between $\rho(x)$ and the EOB potentials $a(x)$ and
$\bar{d}(x)$ on the one hand, with the expressions of
Ref.~\cite{Le.15} for $a(x)$ and $\bar{d}(x)$ in terms of $z_{(1)}(x)$
and $\av{z}_{(1)}^{e^2}(x)$ on the other hand. As an additional check
of Eq.~\eqref{enfin}, \vspace{-0.1cm} we shall also consider the
behaviour of the functions $z_{(1)}(p)$ and \vspace{-0.05cm}
$\av{z}^{e^2}_{(1)}(p)$ in the weak-field limit $p \to +\infty$, and
verify that we recover the known large-$p$ behaviour for $K_{(1)}(p)$,
known from the 4PN calculations of Ref.~\cite{Da.al.16}.

The gauge-invariant relation $\av{z}(\hat{n},\hat{\omega})$ has been
computed for generic orbits, up to 3PN order, for any mass ratio
\cite{Ak.al.15}.\footnote{In App.~\ref{appC} we use the first law
  \eqref{firstlaw} to compute the redshift up to 4PN order, for
  circular orbits only.} From this it is straighforward to derive the
3PN expansions of the $\calO(\nu)$ contributions $z_{(1)}(p)$ and
$\av{z}^{e^2}_{(1)}(p)$ to $\av{z}(\hat{n},\hat{\omega})$. On the
other hand, the application of analytical techniques for linear black
hole perturbations has given access to high-order PN expansions for
these functions. In particular, the contribution
$\av{z}^{e^2}_{(1)}(p)$ has been computed up to 4PN order in
Ref.~\cite{Ho.al.16}, and up to 9.5PN order in
Ref.~\cite{Bi.al2.16}. Combining those results, we find the
4PN-accurate formulas
\begin{subequations}\label{expansions}
\begin{align}
z_{(1)}(p) &= \frac{1}{p} - \frac{1}{p^2} - \frac{1}{p^3} + \left(
\frac{76}{3} - \frac{41}{32} \pi^2 \right) \frac{1}{p^4} \\ &+ \left(
- \frac{658}{15} + \frac{1291}{512} \pi^2 + \frac{128}{5}
\gamma_\text{E} + \frac{256}{5} \ln{2} - \frac{64}{5} \ln{p} \right)
\frac{1}{p^5} + o(p^{-5}) \, ,
\nonumber\\ \frac{\av{z}^{e^2}_{(1)}(p)}{2} &= - \frac{1}{p} +
\frac{2}{p^2} + \frac{5}{p^3} + \left( \frac{23}{3} + \frac{41}{32}
\pi^2 \right) \frac{1}{p^4} \\ &+ \left( \frac{10151}{45} -
\frac{53281}{3072} \pi^2 + \frac{592}{15} \gamma_\text{E} -
\frac{3248}{15} \ln{2} + \frac{1458}{5} \ln{3} - \frac{296}{15} \ln{p}
\right) \frac{1}{p^5} + o(p^{-5}) \,,\nonumber
	\end{align}
\end{subequations}
where $\gamma_\text{E}$ is Euler's constant. Notice the logarithmic
running appearing at 4PN order, related to the occurence of
gravitational-wave tails.

However, the expansions \eqref{expansions} cannot be right away
substituted into the formulas \eqref{enfin} and \eqref{rho}. Indeed,
the former results were derived while normalizing the frequencies
using the black hole mass $m_2$, but the latter results were derived
while normalizing the frequencies using the total mass $m$. Hence, we
first need to account for the correction originating from the
substitutions $\hat{\omega} = m_2\omega + \nu \, m_2\omega +
\mathcal{O}(\nu^2)$ and $\hat{n} = m_2 n + \nu \, m_2 n +
\mathcal{O}(\nu^2)$ in $\av{z}_{(0)}(\hat{\omega},\hat{n})$, which is
simply given by
\beq\label{Dz} - \hat{\omega} \, \frac{\partial \av{z}_{(0)}}{\partial
  \hat{\omega}} - \hat{n} \, \frac{\partial \av{z}_{(0)}}{\partial
  \hat{n}} = \frac{1}{\sqrt{p(p-3)}} \left( 1 -
\frac{2p^3-25p^2+92p-102}{2(p-2)(p-3)(p-6)} \, e^2 + \mathcal{O}(e^4)
\right) , \eeq
where we used Eqs.~\eqref{omega_ep}, \eqref{n_ep}, \eqref{zoom0} and
\eqref{zoom1} to evaluate this expression in the small-eccentricity
limit. Then, adding the 4PN expansion of the correction term
\eqref{Dz} to the 4PN expansions \eqref{expansions}, and substituting
the results in Eqs.~\eqref{enfin} and \eqref{rho}, we obtain the 4PN
expansions of the $\calO(\nu)$ contributions to $K$ and $W = 1/K^2$ as
\begin{subequations}
	\begin{align}
K_{(1)}(p) &= - \frac{7}{p^2} + \left( - \frac{649}{4} +
\frac{123}{32} \pi^2 \right) \frac{1}{p^3} \label{K1} \\ &+ \left( -
\frac{275941}{360} + \frac{48007}{3072} \pi^2 - \frac{592}{15} \ln{2}
- \frac{1458}{5} \ln{3} - \frac{2512}{15} \gamma_\text{E} +
\frac{1256}{15} \ln{p} \right) \frac{1}{p^4} + o(p^{-4}) \, ,
\nonumber \\ \rho(x) &= 14 x^2 + \left( \frac{397}{2} - \frac{123}{16}
\pi^2 \right) x^3 \\ &+ \left( - \frac{215729}{180} +
\frac{58265}{1536} \pi^2 + \frac{1184}{15} \ln{2} + \frac{2916}{5}
\ln{3} + \frac{5024}{15} \gamma_\text{E} + \frac{2512}{15} \ln{x}
\right) x^4 + o(x^4) \, . \nonumber
	\end{align}
\end{subequations}
This last result is in full agreement with the 4PN expansion of the
function $\rho(x)$, as computed up to 9.5PN order using analytic GSF
methods \cite{Bi.al2.16}.

In order to compare the formula \eqref{K1} to the known 4PN result for
$K(\omega)$, one needs to add the contribution from the zero-th order
term in Eq.~\eqref{Kbar}, which can easily be computed by taking the
ratio of Eqs.~\eqref{omega_ep} and~\eqref{n_ep} in the
zero-eccentricity limit, namely
\beq\label{K0} K_{(0)}(p) = \sqrt{\frac{p}{p-6}} = 1 + \frac{3}{p} +
\frac{27}{2p^2} + \frac{135}{2p^3} + \frac{2835}{8p^4} + o(p^{-4}) \,.
\eeq
Expressing the total periastron advance \eqref{Kbar} in terms of the
frequency-related PN parameter $x \equiv \hat{\omega}^{2/3} = p^{-1} +
\calO(\nu)$, rather than the semi-latus rectum $p$, we find that
\eqref{K1} and \eqref{K0} combine to give
\begin{align}
	K(x) &= 1 + 3x + \left( \frac{27}{2} - 7 \nu \right) x^2 +
        \left( \frac{135}{2} + \biggl[ - \frac{649}{4} +
          \frac{123}{32} \pi^2 \biggr] \nu \right) x^3
        \nonumber \\ &+ \biggl( \frac{2835}{8} + \biggl[
          - \frac{275941}{360} + \frac{48007}{3072} \pi^2 -
          \frac{592}{15} \ln{2} - \frac{1458}{5} \ln{3} -
          \frac{2512}{15} \gamma_\text{E} - \frac{1256}{15} \ln{x}
          \biggr] \nu \biggr) \, x^4 \nonumber \\ &+ o(\nu,x^4) \, .
\end{align}
Up to uncontroled terms $\calO(\nu^2)$ and $\calO(\nu^3)$, this result
is in entire agreement with the known 4PN result, as derived for any
mass ratio in the canonical ADM framework \cite{Da.al.16} and in the
harmonic-coordinates Fokker Lagrangian approach \cite{Be.al.17,Be.al2.17}.

Finally, let us check that the binary's binding energy for circular
orbits at the 4PN order is correctly recovered by the same method. For
general orbits, the rescaled binding energy
$\hat{E}\equiv(E-m)/(m\nu)$ is expressed as a function of the
dimensionless frequencies $\hat{\omega}$ and $\hat{n}$. In the small
mass-ratio limit we have $\hat{E}=E_{(0)}+\nu
E_{(1)}+\mathcal{O}(\nu^2)$ where, as a consequence of the first law
(see Eqs.~(5.8a) and (5.9a) in Ref.~\cite{Le.15}),
\begin{subequations}\label{E01}
\begin{align}
E_{(0)}(\hat{\omega},\hat{n}) &= \av{z}_{(0)} - \hat{\omega} \,
\frac{\partial \av{z}_{(0)}}{\partial \hat{\omega}} - \hat{n} \,
\frac{\partial \av{z}_{(0)}}{\partial \hat{n}} - 1\, ,
\\ E_{(1)}(\hat{\omega},\hat{n}) &= \frac{1}{2} \biggl( \av{z}_{(1)} +
2 E_{(0)} - \hat{\omega} \, \frac{\partial \av{z}_{(1)}}{\partial
  \hat{\omega}} - \hat{n} \, \frac{\partial \av{z}_{(1)}}{\partial
  \hat{n}} \nonumber\\ &\qquad\qquad\qquad + \hat{\omega}^2 \,
\frac{\partial^2 \av{z}_{(0)}}{\partial \hat{\omega}^2} + 2
\hat{\omega} \hat{n} \, \frac{\partial^2 \av{z}_{(0)}}{\partial
  \hat{\omega} \partial \hat{n}} + \hat{n}^2 \, \frac{\partial^2
  \av{z}_{(0)}}{\partial \hat{n}^2} \biggr) \,.\label{E01b}
	\end{align}
\end{subequations}
As before we parameterize each of these quantities by means of the
Schwarzschild semi-latus rectum $p$ and eccentricity $e$, rather than
by $\hat{\omega}$ and $\hat{n}$. Thanks to our previous computation of
the periastron advance for circular orbits, \vspace{-0.05cm} it is
simple to deduce from \eqref{E01} the circular-orbit limit of the
energy, say $\hat{E}^\text{circ} = E_{(0)}^\text{circ}(\hat{\omega}) +
\nu E_{(1)}^\text{circ}(\hat{\omega})+\mathcal{O}(\nu^2)$. Indeed,
while $E_{(0)}^\text{circ}$ is obviously given by $E_{(0)}$ for
circular orbits (\textit{i.e.}, by taking $e\to 0$ and then changing
$p^{-1}=x$), the GSF contribution $E_{(1)}^\text{circ}$ is not
directly given by the circular limit of~\eqref{E01b}. Rather, it
receives an additional contribution, explicitly reading
\beq\label{E(1)circ} E_{(1)}^\text{circ} = E_{(1)} - \hat{\omega}
\,\frac{K_{(1)}}{K^2_{(0)}} \,\frac{\partial E_{(0)}}{\partial
  \hat{n}}\,, \eeq
where the right-hand side is evaluated for $e=0$ and $p=x^{-1}$. By
this method we recover the known 4PN results for the GSF limit of the
circular binding energy, namely
\cite{Le.al.12,BiDa.13,Da.al.14,Be.al.17}
\begin{align}\label{Ecirc}
\!\!\! E(x) &= m -\frac{m\nu x}{2} \biggl\{ 1 + \left( - \frac{3}{4} -
\frac{\nu}{12} \right) x + \left( - \frac{27}{8} + \frac{19}{8} \nu
\right) x^2 + \left( - \frac{675}{64} + \biggl[ \frac{34445}{576} -
  \frac{205}{96} \pi^2 \biggr] \nu \right) x^3 \nonumber \\ &+ \left(
- \frac{3969}{128} + \left[-\frac{123671}{5760}+\frac{9037}{1536}\pi^2
  + \frac{896}{15}\gamma_\text{E}+ \frac{448}{15} \ln(16 x)\right]\nu
\right) x^4 + o(\nu,x^4)\biggr\} \,.
\end{align}
The angular momentum $L(x)$ can be computed in the same way.  In that
case, the relevant formulas for the rescaled momentum $\hat{L}\equiv
L/(m^2\nu)$ are Eqs.~(5.8b) and (5.9b) in Ref.~\cite{Le.15}, and we
add a correction term similar to the one in \eqref{E(1)circ}. The
result reads
\begin{align}\label{Lcirc}
L(x) &= \frac{m^2\nu}{\sqrt{x}} \biggl\{ 1 + \left(
  \frac{3}{2} + \frac{\nu}{6} \right) x + \left( \frac{27}{8} -
  \frac{19}{8} \nu \right) x^2 + \left( \frac{135}{16} + \biggl[ -
    \frac{6889}{144} + \frac{41}{24} \pi^2 \biggr] \nu \right) x^3
\nonumber \\ &+ \left( \frac{2835}{128} + \left[
    \frac{98869}{5760}-\frac{6455}{1536}\pi^2 -
    \frac{128}{3}\gamma_\text{E} - \frac{64}{3} \ln(16 x)\right]\nu
  \right) x^4 + o(\nu,x^4)\biggr\} \,.
\end{align}
Of course, we may explicitly check that $\ud E/\ud x = \omega \,\ud
L/\ud x$ at fixed masses.

\acknowledgments

LB thanks Bernard Whiting for discussions on gravitational self-force
numerical results. ALT acknowledges financial support through a Marie
Curie FP7 Integration Grant within the 7th European Union Framework
Programme (PCIG13-GA-2013-630210).

\appendix

\section{Fourier series and long-time average}
\label{appA}

The components of the mass quadrupole moment $I_{ij}$ of generic
elliptic orbits at Newtonian order, in the center-of-mass frame, are
decomposed into the discrete Fourier series
\beq\label{Iij_Fourier} I_{ij}(t) =
\sum_{p=-\infty}^{+\infty}\,\mathop{{\mathcal{I}}}_{p}{}_{\!\!ij}\,\ue^{\ui
  p\ell} \,,\eeq
where $\ell=n(t-t_0)$ is the mean anomaly, with $n=2\pi/P$ the
frequency associated to the period $P$ of the orbital motion, and
$t_0$ is some instant of passage at periastron. The Fourier
coefficients ${}_p\mathcal{I}_{ij}$ depend on $n$ and the orbit's
eccentricity $e$, and are fully available as closed-form combinations
of Bessel functions in App.~B of \cite{Be.al.17} and App.~A of
\cite{Ar.al2.08}. Averaging over one orbital period, we get
\beq\label{avIij} \langle I_{ij} \rangle = \int_0^{2\pi} \frac{\ud
  \ell}{2\pi}\,I_{ij}(\ell) =
\mathop{{\mathcal{I}}}_{0}{}_{\!\!ij}\,.\eeq
However, in this paper it is important to define the time average of
a function $f(t)$ in a more general manner, when the function is not
necessarily periodic, by 
\beq\label{longtermaverage2} \langle f \rangle \equiv \lim_{T\to
  +\infty}\frac{1}{2T}\int_{-T}^{T}\ud t\,f(t)\,. \eeq
Such a long-time average coincides with the usual average for periodic
functions. An important property of the long-time average
\eqref{longtermaverage2} is that it implies $\langle \dot{f}
\rangle=0$ for any function $f$ that remains bounded when
$t\to\pm\infty$.

Most relevant quantities can be evaluated explicitly by inserting the
Fourier series \eqref{Iij_Fourier}. For instance, the tail
factor \eqref{Tijs} reads as
\begin{equation}\label{tailfactorFourier}
\mathcal{T}_{ij}^{(s)} = -2 \sum_{p=-\infty}^{+\infty} (\ui
p\,n)^s\,\mathop{{\mathcal{I}}}_{p}{}_{\!\!ij} \Bigl(\ln\left(2\vert
p\vert n r\right) + \gamma_\text{E} \Bigr)\ue^{\ui p \ell}\,,
\end{equation}
where we recall that the separation $r$ between the particles has been
used as the Hadamard Pf scale. The quantity $Q_H$ that was defined in
Sec.~\ref{sec:nonlocal} to be the right-hand side of Eq.~\eqref{dotH},
and which is such that $\dot{H}=Q_H$, can be obtained by a
straightforward computation as the following (double) Fourier
series\footnote{We observe that here the Hadamard partie finie scale
  $r$ has cancelled out.}
\beq\label{QH_Fourier} Q_H = - \frac{m}{5} \,n^7 \sum_{p+q \not=
  0}\ui\mathop{{\mathcal{I}}}_{p}{}_{\!\!ij}
\mathop{{\mathcal{I}}}_{q}{}_{\!\!ij} \,p^3q^3(p-q)
\,\ln\left|\frac{p}{q}\right|\,\ue^{\ui(p+q)\ell}\,. \eeq
Since $Q_H$ contains only modes with $p+q \not= 0$, it averages to
zero: $\langle Q_H\rangle=0$. The oscillatory correction term $\Delta
H^\text{AC}$ in the conserved energy, as defined by Eqs.~\eqref{E}
and~\eqref{HAC}, can be obtained directly by integrating term by term
Eq.~\eqref{QH_Fourier}. Indeed, it is necessary and sufficient
to discard any integration constant so that $\langle\Delta
H^\text{AC}\rangle=0$, and we obtain
\beq\label{HAC_Fourier} \Delta H^\text{AC} = \frac{m}{5} \, n^6
\sum_{p+q \not= 0}\,\mathop{{\mathcal{I}}}_{p}{}_{\!\!ij}
\mathop{{\mathcal{I}}}_{q}{}_{\!\!ij} \,\frac{p^3q^3(p-q)}{p+q}
\ln\left|\frac{p}{q}\right|\,\ue^{\ui(p+q)\ell}\,. \eeq
On the other hand, the actual integration constant which is to be
added to get the conserved energy $E$ requires a separate
analysis, which was performed in Ref.~\cite{Be.al.17}. The result is
the DC term given in Eq.~\eqref{HDC}, which is proportional to the
total averaged gravitational-wave energy flux.

Next, we present some formulas concerning the angular momentum,
and notably the AC correction term $\Delta p_\varphi^\text{AC}$
therein, which as we have seen enters into the modified radial action
integral intervening into the first law; see Eq.~\eqref{calR}. The
Hamiltonian equation for $p_\varphi$ was given in
Eq.~\eqref{dotpphi}. With spatial coordinates $(x,y,z)$ adapted to the
orbital motion into the plane $(x,y)$, \textit{i.e.} such that the
moving triad in the orbital plane reads $\bo{n} = (\cos\varphi,
\sin\varphi, 0)$, $\bm{\lambda} = (-\sin\varphi, \cos\varphi, 0)$ and
$\bm{\ell}=\bo{n}\times\bm{\lambda} = (0,0,1)$, we have
\beq\label{dIdphi} \frac{1}{2}\frac{\partial
  \hat{I}_{ij}^{(3)}}{\partial \varphi} =
\ell^k\epsilon_{kl \langle i}\hat{I}_{j \rangle l}^{(3)}\,,\eeq
where the brackets around indices denote the STF projection.
This equation can be checked for instance using the explicit
expression~\eqref{Iij3}. Hence we can readily express the right-hand
side of the angular momentum equation~\eqref{dotpphi} as the following
double Fourier series,
\beq\label{Qpphi_Fourier} Q_{p_\varphi} = - \, \frac{4m}{5} \,n^6
\sum_{p+q \not= 0}\,\mathop{\mathcal{K}}_{p,q}
\,p^3q^3\,\ln\left|\frac{p}{q}\right|\,\ue^{\ui(p+q)\ell}\,.\eeq
It involves only non-zero modes $p+q \not= 0$, and we have defined
\beq\mathop{\mathcal{K}}_{p,q} \equiv
\ell^i\epsilon_{ijk}\mathop{{\mathcal{I}}}_{p}{}_{\!\!jl}
\mathop{{\mathcal{I}}}_{q}{}_{\!\!kl} =
\bigl(\mathop{{\mathcal{I}}}_{p}{}_{\!\!xx}-
\mathop{{\mathcal{I}}}_{p}{}_{\!\!yy}\bigr)\mathop{{\mathcal{I}}}_{q}{}_{\!\!xy}
- \mathop{{\mathcal{I}}}_{p}{}_{\!\!xy}
\bigl(\mathop{{\mathcal{I}}}_{q}{}_{\!\!xx}-
\mathop{{\mathcal{I}}}_{q}{}_{\!\!yy}\bigr)\,.\eeq
By integrating term by term the Fourier series~\eqref{Qpphi_Fourier},
and ignoring any additive integration constant, we obtain directly the
AC correction piece in the conserved angular momentum as defined
by~\eqref{pphiAC}:
\beq\label{ACfourier} \Delta p_\varphi^\text{AC} = - \frac{4m}{5}
\,n^5 \sum_{p+q \not= 0}\,\ui\!\mathop{\mathcal{K}}_{p,q}
\frac{p^3q^3}{p+q}
\,\ln\left|\frac{p}{q}\right|\,\ue^{\ui(p+q)\ell}\,,\eeq
which is such that $\langle\Delta p_\varphi^\text{AC}\rangle=0$. On
the other hand, the obtention of the constant DC piece is less
trivial~\cite{Be.al.17} and the result has been given in
Eq.~\eqref{pphiDC}.

Finally, we want to control the extra term that was found in the
effective action integral appearing into the first law. According to
\eqref{calR} we have $\mathscr{R} = R + 2m (\mathscr{G} -
  \mathscr{F}/\omega) - I$ with
\beq\label{I} I = \frac{1}{2\pi} \oint \Delta p_\varphi^\text{AC}\,\ud
\varphi = \frac{1}{n}\,\langle \dot\varphi\Delta
p_\varphi^\text{AC}\rangle\,. \eeq
The Fourier transform of the ``instantaneous'' frequency $\dot\varphi$
is known to the Newtonian order, which is sufficient here
since~\eqref{I} is a small 4PN quantity. We have (see
\textit{e.g.}~\cite{Dur})
\beq\label{phipoint} \dot{\varphi} = n \biggl[ 1 + 2
  \sum_{k=1}^{+\infty} \alpha_k \cos(k\ell)\biggr]\,, \eeq
where the coefficients read [with $f\equiv(1-\sqrt{1-e^2})/e$ and
  $J_k$ being the usual Bessel function]
\beq\label{alphak} \alpha_k = J_k(k e) + \sum_{s=1}^{+\infty}
f^s\Bigl[J_{k-s}(k e)+J_{k+s}(k e)\Bigr]\,. \eeq
Therefore by inserting into \eqref{I} both the Fourier series for
the instantaneous frequency~\eqref{phipoint} and that for the AC
correction term in the angular momentum~\eqref{ACfourier}, we obtain
the result
\beq\label{Iresult} I = - \frac{4m}{5} \,n^5 \sum_{p+q \not=
  0}\,\ui\!\mathop{\mathcal{K}}_{p,q}
\,\alpha_{\left|p+q\right|}\,\frac{p^3q^3}{p+q}
\,\ln\left|\frac{p}{q}\right| .\eeq
(Since $\overline{\mathcal{K}}_{p,q} = \mathcal{K}_{-p,-q}$ we can
check that $I$ is real.) For circular orbits one must have $p=\pm
  \,2$ and $q=\pm \,2$, such that $I=0$ in that case.

\section{Small-eccentricity limit}
\label{appB}

In this appendix, we collect some results that were used in
Sec.~\ref{sec:K-z}, for a test mass orbiting around a Schwarzschild
black hole of mass $M$, for nearly circular orbits. Hereafter, we omit
the subscript $(0)$, but all of the formulas below hold only in the
test-mass limit. Recall that the frequencies $n$ and $\omega$ are
normalized using the total mass, which here reduces to the black hole
mass, \textit{i.e.} $(\hat{\omega},\hat{n}) = (M\omega,Mn)$ in this
appendix.

Instead of parameterizing the bound timelike geodesic of the test
particle by means of the frequencies $\hat{\omega}$ and $\hat{n}$, or
alternatively by means of the conserved specific energy $\mathcal{E}$
and specific angular momentum $\mathcal{L}$, we shall use the
convenient ``semi-latus rectum'' $p$ and ``eccentricity'' $e$, defined
such that \cite{Cu.al.94}
\beq\label{EL-pe} \mathcal{E} = \left[
  \frac{(p-2-2e)(p-2+2e)}{p(p-3-e^2)} \right]^{1/2} \,, \qquad
\mathcal{L} = \frac{pM}{\sqrt{p-3-e^2}} \, .  \eeq
Following \cite{Da.61}, we parameterize the particle's radial motion
(in Schwarzschild coordinates) using the ``relativistic anomaly''
$\chi$ as
\beq\label{r-chi} r(\chi) = \frac{p M}{1 + e \cos{\chi}} \, , \eeq
where $\chi = 0$ and $\chi = \pi$ correspond to the periastron and the
apastron passages, respectively.  In terms of the orbital parameters
$p$ and $e$, we have the usual Newtonian-looking expressions $p = 2
r_+ r_- / [M(r_+ + r_-)]$ and $e = (r_+ - r_-) / (r_+ + r_-)$.

Combining Eqs.~\eqref{EL-pe} and~\eqref{r-chi} with the well-known
(first integral form of the) geodesic equations of motion for a test
particle in Schwarzschild spacetime, the coordinate time period of the
radial motion, $P$, the corresponding proper time period, $T$, as well
as the accumulated azimuthal angle per radial period, $\Phi$, are
given by the definite integrals \cite{Cu.al.94,BaSa.10,Ak.al.15}
\begin{subequations}\label{integrals}
\begin{align}
P(p,e) &= \int_0^{2\pi} \! \frac{\ud t}{\ud \chi} \, \ud \chi =
\int_0^{2\pi} \! \frac{M p^2
  \sqrt{(p-2-2e)(p-2+2e)}}{(p-2-2e\cos\chi)(1+e\cos\chi)^2
  \sqrt{p-6-2e\cos\chi}} \, \ud \chi \, , \\ T(p,e) &= \int_0^{2\pi}
\! \frac{\ud \tau}{\ud \chi} \, \ud \chi = \int_0^{2\pi} \! \frac{M
  p^{3/2}}{(1+e \cos\chi)^2} \sqrt{\frac{p-3-e^2}{p-6-2e\cos\chi}} \,
\ud \chi \, , \\ \Phi(p,e) &= \int_0^{2\pi} \! \frac{\ud \varphi}{\ud
  \chi} \, \ud \chi = 4 \sqrt{\frac{p}{p-6+2e}} \, \text{ellipK}
\left( \frac{4e}{p-6+2e} \right) ,
	\end{align}
\end{subequations}
where $\text{ellipK}(k) \equiv \int_0^{\pi/2} (1 - k \sin^2
\theta)^{-1/2} \, \ud \theta$ is the complete elliptic integral of the
first kind. Then, the radial frequency $n$, the averaged azimuthal
frequency $\omega$, and the averaged redshift variable $\av{z}$ are
defined as
\beq n \equiv \frac{2\pi}{P} \, , \qquad \omega \equiv \frac{\Phi}{P}
\, , \qquad \av{z} \equiv \frac{T}{P} \, .  \eeq

No closed form expressions for $n(p,e)$, $\omega(p,e)$ and
$\av{z}(p,e)$ are known. Still, the definite
integrals \eqref{integrals} can be computed in the small-eccentricity
limit $e \ll 1$, yielding the following Taylor series expansions:
\begin{subequations}\label{youpla}
\begin{align}
\hat{\omega}(p,e) &= \frac{1}{p^{3/2}} - \frac{3}{2} \,
\frac{p^2-10p+22}{p^{3/2} (p-2) (p-6)} \, e^2 + \calO(e^4) \,
, \label{omega_ep} \\ 
\hat{n}(p,e) &= \frac{\sqrt{p-6}}{p^2} - \frac{3}{4} \,
\frac{2p^3-32p^2+165p-266}{p^2 (p-2) (p-6)^{3/2}} \, e^2 + \calO(e^4)
\, , \label{n_ep} \\
\av{z}(p,e) &= \sqrt{\frac{p-3}{p}} + \frac{3}{2} \,
\frac{p^2-10p+22}{\sqrt{p(p-3)}(p-2)(p-6)} \, e^2 + \calO(e^4)\,.
\end{align}
\end{subequations}
Here, we gave the results up to $\calO(e^2)$ only, because the
formulas become too cumbersome at higher orders. However the
expansions \eqref{youpla} can in principle be computed up to
arbitrarily high orders in powers of $e^2$. Then, the partial
derivatives of the dimensionless frequencies $\hat{\omega}$, $\hat{n}$
and $\av{z}$ with respect to the orbital parameters $p$ and $e$ read
as
\begin{subequations}\label{youpi}
\begin{align}
\left( \frac{\partial \hat{\omega}}{\partial p} \right)_{\!e} &= -
\frac{3}{2p^{5/2}} + \calO(e^2) \, , \label{youpi1} \\ 
\left( \frac{\partial \hat{\omega}}{\partial e} \right)_{\!p} &= -
\frac{3(p^2-10p+22)}{p^{3/2} (p-2) (p-6)} \, e + \calO(e^3) \,
, \label{youpi2} \\
\left( \frac{\partial \hat{n}}{\partial p}
\right)_{\!e} &= - \frac{3}{2} \, \frac{p-8}{p^3 \sqrt{p-6}} +
\calO(e^2) \, , \label{youpi3} \\
\left( \frac{\partial \hat{n}}{\partial e} \right)_{\!p} &= -
\frac{3}{2} \, \frac{2p^3-32p^2+165p-266}{p^2 (p-2) (p-6)^{3/2}} \, e
+ \calO(e^3) \, , \label{youpi4} \\
\left( \frac{\partial \av{z}}{\partial p} \right)_{\!e} &= \frac{3}{2
  p^{3/2} \sqrt{p-3}} + \calO(e^2) \, , \label{youpi5} \\
\left( \frac{\partial \av{z}}{\partial e} \right)_{\!p} &=
\frac{3(p^2-10p+22)}{\sqrt{p(p-3)} (p-2) (p-6)} \, e + \calO(e^3) \,
. \label{youpi6}
\end{align}
\end{subequations}
From these expressions, one can easily compute the determinant of the
matrix transformation from $(p,e)$ to $(\hat{\omega},\hat{n})$, namely
\beq\label{det} D \equiv \left| \frac{\partial
  (\hat{\omega},\hat{n})}{\partial(p,e)} \right| = \frac{9}{4} \,
\frac{4p^2-39p+86}{p^{9/2}(p-2)(p-6)^{3/2}} \, e + \calO(e^3) \, .
\eeq
Combining the expansions \eqref{youpi1}--\eqref{youpi4} and \eqref{det},
we get the following expressions for partial derivatives that appear,
among others, in Eq.~\eqref{pouf}:
\begin{subequations}\label{tac}
\begin{align}
\left( \frac{\partial p}{\partial \hat{n}} \right)_{\!\hat{\omega}} &=
- \frac{1}{D} \left( \frac{\partial \hat{\omega}}{\partial e} \right)_{\!p} =
\frac{4}{3} \, p^3 \sqrt{p-6} \, \frac{p^2 - 10 p + 22}{4p^2 - 39p +
  86} + \calO(e^2) \, , \label{tac1} \\ 
\left( \frac{\partial e}{\partial \hat{n}} \right)_{\!\hat{\omega}} &=
+ \frac{1}{D} \left( \frac{\partial \hat{\omega}}{\partial p} \right)_{\!e} = -
\frac{2}{3} \, \frac{p^2 (p-2) (p-6)^{3/2}}{4p^2 - 39p +86} \,
\frac{1}{e} + \calO(e) \, , \label{tac2} \\
\left( \frac{\partial p}{\partial \hat{\omega}} \right)_{\!\hat{n}} &=
+ \frac{1}{D} \left( \frac{\partial \hat{n}}{\partial e} \right)_{\!p}
= - \frac{2}{3} \, p^{5/2} \, \frac{2p^3 - 32p^2 + 165p - 266}{4p^2 -
  39p + 86} + \calO(e^2) \, , \label{tac3} \\
\left( \frac{\partial e}{\partial \hat{\omega}} \right)_{\!\hat{n}} &=
- \frac{1}{D} \left( \frac{\partial \hat{n}}{\partial p} \right)_{\!e}
= \frac{2}{3} \, p^{3/2} \, \frac{(p-2)(p-6)(p-8)}{4p^2 - 39p +86} \,
\frac{1}{e} + \calO(e) \, . \label{tac4}
	\end{align}
\end{subequations}
Finally, combining Eqs.~\eqref{youpi5}, \eqref{youpi6} and
\eqref{tac}, and using the chain rule from $(p,e)$ to
$(\hat{\omega},\hat{n})$, we obtain the following expressions for the
frequency derivatives of the average redshift that appear in
Eqs.~\eqref{K_GSF}, \eqref{Dz}, \eqref{E01} and \eqref{E(1)circ}:
\begin{subequations}\label{trucs}
\begin{align}
\frac{\partial \av{z}}{\partial \hat{n}} &= - \frac{1}{2}
\sqrt{\frac{p-6}{p-3}} \frac{p^{3/2}}{p-2} \, e^2 + \calO(e^4) \,
, \label{zoom0} \\
\frac{\partial \av{z}}{\partial \hat{\omega}} &= -
\frac{p}{\sqrt{p-3}} \left( 1 + \frac{e^2}{2(p-3)} + \calO(e^4)
\right) , \label{zoom1} \\
\frac{\partial^2 \av{z}}{\partial \hat{n}^2} &=
\frac{2}{3} \, \frac{p^{7/2} (p-6)^2}{\sqrt{p-3} (4p^2 - 39p + 86)} +
\calO(e^2) \, , \label{zoom2} \\
\frac{\partial^2 \av{z}}{\partial \hat{\omega} \partial \hat{n}} &= -
\frac{2}{3} \, \frac{p^3 (p-6)^{3/2} (p-8)}{\sqrt{p-3} (4p^2 - 39p +
  86)} + \calO(e^2) \, , \label{zoom3} \\
\frac{\partial^2 \av{z}}{\partial \hat{\omega}^2} &=
  \frac{p^{5/2} (p-6) (2p^3-34p^2+185p-298)}{3(p-3)^{3/2} (4p^2 - 39p
    + 86)} + \calO(e^2) \, . \label{zoom4}
	\end{align}
\end{subequations}
The calculation of these partial derivatives requires the control of
$\av{z}(p,e)$ up to $\calO(e^4)$, and that of all derived quantities
at the same relative order in $e^2$. Note that the first derivative
\eqref{zoom0}, which is $\calO(e^2)$, does not contribute to the final
circular-orbit result in Eq.~\eqref{enfin}.

\section{Redshift for circular orbits}
\label{appC}

In this section, we derive the 4PN expressions for the particles'
redshifts in the particular case of circular orbits. For such orbits,
$\mathscr{R} = 0$ and the first law \eqref{firstlaw} implies
\beq \frac{\partial E}{\partial \omega} \bigg|_{m_a} = \omega \,
\frac{\partial L}{\partial \omega} \bigg|_{m_a} \,.  \eeq
Moreover, by considering variations with respect to the particles'
masses $m_a$ at fixed circular-orbit frequency $\omega$, the first law
\eqref{firstlaw} yields the following expression for the constant
redshift $z_a \equiv \av{z_a}$ of each particle:
\beq\label{z_a} z_a = \frac{\partial E}{\partial m_a} \bigg|_\omega -
\omega \, \frac{\partial L}{\partial m_a} \bigg|_\omega =
\frac{\partial \mathcal{M}}{\partial m_a} \bigg|_\omega , \eeq
where we introduced $\mathcal{M} \equiv E - \omega L$, heuristically
the binary's energy in a co-rotating frame.
Now, the expressions for the conserved circular-orbit
energy $E(\omega)$ and the angular momentum $L(\omega$) were recently
derived up to 4PN order \cite{Da.al.14,Be.al.16}. By substituting for
Eqs.~(5.4b) and (5.5) of Ref.~\cite{Da.al.14} into Eq.~\eqref{z_a}, we
obtain the 4PN-accurate expression for the constant redshift of
particle $1$ as
\begin{align}\label{z1NS}
z_1 &= 1 + \left( - \frac{3}{4} - \frac{3}{4} \Delta + \frac{\nu}{2}
\right) x + \left( - \frac{9}{16} - \frac{9}{16} \Delta -
\frac{\nu}{2} - \frac{1}{8} \Delta \, \nu + \frac{5}{24} \nu^2 \right)
x^2 \nonumber \\ &\qquad\! + \left( - \frac{27}{32} - \frac{27}{32}
\Delta - \frac{\nu}{2} + \frac{19}{16} \Delta \, \nu - \frac{39}{32}
\nu^2 - \frac{1}{32} \Delta \, \nu^2 + \frac{\nu^3}{16} \right) x^3
\nonumber \\ &\qquad\! + \left( - \, \frac{405}{256} - \frac{405}{256}
\Delta + \left[ \frac{38}{3} - \frac{41}{64} \pi^2 \right] \nu +
\left[ \frac{6889}{384} - \frac{41}{64} \pi^2 \right] \Delta \, \nu
\right. \nonumber \\ &\qquad\qquad\!\! \left. + \left[ -
  \frac{3863}{576} + \frac{41}{192} \pi^2 \right] \nu^2 -
\frac{93}{128} \Delta \, \nu^2 + \frac{973}{864} \nu^3 -
\frac{7}{1728} \Delta \, \nu^3 + \frac{91}{10368} \nu^4 \right) x^4
\nonumber \\ &\qquad\! + \left( - \, \frac{1701}{512} - \,
\frac{1701}{512} \Delta + \left[ - \frac{329}{15} + \frac{1291}{1024}
  \pi^2 + \frac{64}{5} \gamma_\text{E} + \frac{32}{5} \ln{(16x)}
  \right] \nu \right. \nonumber \\ &\qquad\qquad\!\! + \left[ -
  \frac{24689}{3840} + \frac{1291}{1024} \pi^2 + \frac{64}{5}
  \gamma_\text{E} + \frac{32}{5} \ln{(16x)} \right] \Delta \, \nu +
\left[ - \frac{71207}{1536} + \frac{451}{256} \pi^2 \right] \Delta \,
\nu^2 \nonumber \\ &\qquad\qquad\!\! + \left[ - \frac{1019179}{23040}
  + \frac{6703}{3072} \pi^2 + \frac{64}{15} \gamma_\text{E} +
  \frac{32}{15} \ln{(16x)} \right] \nu^2 + \left[ \frac{356551}{6912}
  - \frac{2255}{1152} \pi^2 \right] \nu^3 \nonumber
\\ &\qquad\qquad\!\! \left. + \,\, \frac{43}{576} \, \Delta \, \nu^3 -
\frac{5621}{41472} \, \nu^4 + \frac{55}{41472} \, \Delta \, \nu^4 -
\frac{187}{62208} \, \nu^5 \right) x^5 + o(x^5) \, ,
\end{align}
where $x \equiv (m \omega)^{2/3}$ is the frequency-related PN
parameter and $\Delta \equiv (m_2 - m_1) / m = \sqrt{1 - 4 \nu}$ the
reduced mass difference. (We assume $m_1 \leqslant m_2$). The
expression for $z_2$ is easily deduced by setting $\Delta \to -\Delta$
in Eq.~\eqref{z1NS}. The expression \eqref{z1NS} is valid for
comparable masses, and in the small mass-ratio limit $\nu\to 0$ we
obtain
\begin{align}\label{z1SF}
z_1 &= 1 + \left( - \frac{3}{2} + 2\nu \right) x + \left( -
\frac{9}{8} + \frac{\nu}{2} \right) x^2 + \left( - \frac{27}{16} +
\frac{19}{8}\nu \right) x^3 + \left( - \frac{405}{128} + \biggl[
  \frac{1621}{48} - \frac{41}{32} \pi^2 \biggr] \nu \right) x^4
\nonumber \\ &+ \left( - \frac{1701}{256} + \left[ -
  \frac{41699}{1920}+\frac{1291}{512}\pi^2 +
  \frac{128}{5}\gamma_\text{E} + \frac{64}{5} \ln(16 x)\right]\nu
\right) x^5 + o(\nu,x^5) \,.
\end{align}

\bibliography{}

\end{document}